\documentclass{article}
\usepackage{authblk}   % 用于作者和机构编号
\usepackage{arxiv}
\usepackage{subcaption}
\usepackage[numbers]{natbib}

\usepackage[utf8]{inputenc} % allow utf-8 input
\usepackage[T1]{fontenc}    % use 8-bit T1 fonts
\usepackage{hyperref}       % hyperlinks
\usepackage{url}            % simple URL typesetting
\usepackage{booktabs}       % professional-quality tables
\usepackage{amsfonts}       % blackboard math symbols
\usepackage{nicefrac}       % compact symbols for 1/2, etc.
\usepackage{microtype}      % microtypography
\usepackage{lipsum}         % Can be removed after putting your text content
\usepackage{graphicx}
\usepackage{doi}

\usepackage{xcolor}
\usepackage{graphicx}
\usepackage{caption}
\usepackage{mlmath}
\usepackage{tikz}
\usetikzlibrary{shapes.geometric, arrows}
\usepackage{cleveref}       % smart cross-referencing

\begin{document}

\title{Neural Network–Based Framework for Passive Intermodulation Cancellation in MIMO Systems}

\newif\ifuniqueAffiliation
% Comment to use multiple affiliations variant of author block 
\uniqueAffiliationtrue

\ifuniqueAffiliation % Standard variant of author block

\author[1]{Xiaolong Li\thanks{\texttt{15369855310@sjtu.edu.cn}}}
\author[1]{Zhi-Qin John Xu\thanks{\texttt{xuzhiqin@sjtu.edu.cn}}}
\author[2]{Peiting You\thanks{\texttt{youpeiting@huawei.com}}}
\author[1]{Yifei Zhu\thanks{\texttt{karinafff@sjtu.edu.cn}}}

\affil[1]{School of Mathematical Sciences, Institute of Natural Sciences, MOE-LSC, Shanghai Jiao Tong University}
\affil[2]{Huawei Technologies Co., Ltd., Shenzhen, Guangdong, China}

\date{} % 如果不想要日期就留空

\else
\fi

% Uncomment to override the `A preprint' in the header
%\renewcommand{\headeright}{Technical Report}
%\renewcommand{\undertitle}{Technical Report}
\renewcommand{\shorttitle}{Neural Network–Based Framework for Passive Intermodulation Cancellation in MIMO Systems}

\maketitle

\begin{abstract}

Passive intermodulation (PIM) has emerged as a critical source of self-interference in modern MIMO-OFDM systems, especially under the stringent requirements of 5G and beyond. Conventional cancellation methods often rely on complex nonlinear models with limited scalability and high computational cost. In this work, we propose a lightweight deep learning framework for PIM cancellation that leverages depthwise separable convolutions and dilated convolutions to efficiently capture nonlinear dependencies across antennas and subcarriers. To further enhance convergence, we adopt a cyclic learning rate schedule and gradient clipping. In a controlled MIMO experimental setup, the method effectively suppresses third-order passive intermodulation (PIM) distortion, achieving up to 29~dB of average power error (APE) with only 11k trainable parameters. These results highlight the potential of compact neural architectures for scalable interference mitigation in future wireless communication systems.
\end{abstract}


\section{Background and Problem Formulation}

\subsection{Background}

With the global deployment of 5G networks, wireless communication has become a critical infrastructure in modern society. Signals in wireless channels propagate through multiple paths, resulting in complex and dynamic fading in the spatial, temporal, and frequency domains. While multipath fading can degrade system performance, it also enables spatial multiplexing, which can significantly increase channel capacity. Consequently, Multiple-Input Multiple-Output (MIMO) technology has emerged as a key approach to exploit spatial resources efficiently \citep{tse2005fundamentals, larsson2014massive}.

MIMO systems deploy multiple antennas at both the transmitter and receiver, allowing simultaneous transmission of multiple data streams. A typical $\mathrm{M} \times \mathrm{N}$ MIMO system has M transmit antennas and N receive antennas. Let $\mathbf{x} = \mathbf{w} s$ denote the transmitted signal using beamforming, where $\mathbf{w}$ is the beamforming vector and $s$ is the transmitted symbol. The corresponding received signal can be expressed as
\[
\mathbf{y} = \mathbf{H} \mathbf{x} + \rho \mathbf{n} = \mathbf{H w} s + \rho \mathbf{n},
\]
where $\mathbf{H} \in \mathcal{C}^{N \times M}$ represents the channel matrix, and $\rho \mathbf{n}$ is additive noise. The channel matrix $\mathbf{H}$ contains $M \times N$ physical channels, typically assumed identically distributed but correlated.

\begin{figure}[h]
    \centering
    \includegraphics[width=0.5\linewidth]{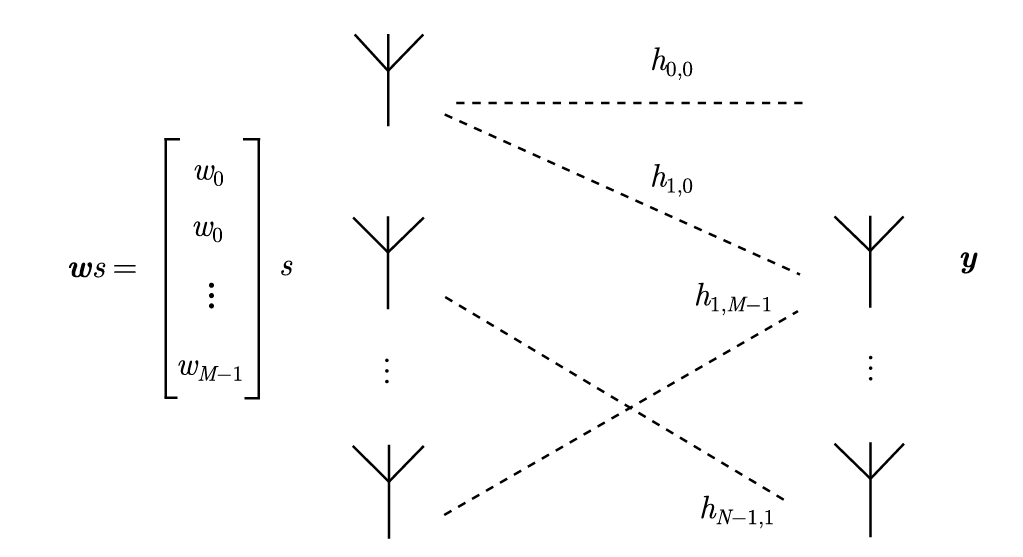}
    \caption{MIMO system}
    \label{fig:MIMO_system}
\end{figure}

To further enhance capacity, 5G New Radio (NR) employs Carrier Aggregation (CA), allowing multiple component carriers (CCs) to be transmitted either continuously or non-continuously within the available spectrum. Frequency Division Duplex (FDD) is also used, where the transmitter and receiver operate simultaneously on different center frequencies. When non-continuously allocated carriers pass through the same nonlinear passive components, new frequency components can be generated, leading to intermodulation products that may fall into the receiver band. When these distortions originate from passive devices, they are referred to as Passive Intermodulation (PIM). PIM can arise from internal sources in the RF chain (e.g., filters, connectors, switches, antennas) or external sources in the radiation field (e.g., metallic structures near the antenna) \citep{johnson2007pim, furuya2018pim}.

\begin{figure}[h]
    \centering
    \includegraphics[width=0.6\linewidth]{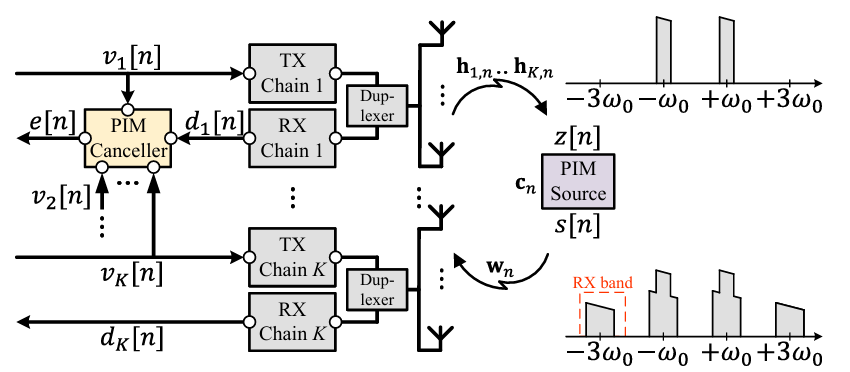}
    \caption{Passive Intermodulation (PIM) in wireless systems}
    \label{fig:PIM}
\end{figure}

As 5G networks continue to expand with more frequency bands and higher transmit power, PIM-induced interference becomes increasingly significant. The interference power can easily exceed the weak received signals, even with high-quality RF components. Therefore, mitigating PIM is essential for reliable communication. Traditional approaches, such as reducing transmit power or improving component quality, often involve trade-offs in cost or performance. An alternative is digital PIM cancellation algorithms, which exploit the deterministic nature of interference to suppress it in the digital front-end of transceiver systems \citep{li2021digital}.

\subsection{Problem Formulation}

Assume a MIMO system where the transmitted signal at the transmitter is denoted as $\mathbf{x} = [x_1, \ldots, x_{T_x}]^T$, and the received signal at the receiver is $\mathbf{y} = [y_1, \ldots, y_{R_x}]^T$. Due to PIM interference, the received signal is distorted into $\tilde{\mathbf{y}} = [\tilde{y}_1, \ldots, \tilde{y}_{R_x}]^T$. Define the PIM interference term as
\[
\mathbf{z} = \tilde{\mathbf{y}} - \mathbf{y},
\]
which can be modeled as a multidimensional function of the transmitted signal, i.e., $\mathbf{z} = f(\mathbf{x})$. Given a limited set of training data $\{\mathbf{x}^{(t)}, \mathbf{y}^{(t)}\}_{t=1}^T$, the goal is to design a deep learning model $f_\theta(\mathbf{x})$ such that
\[
f_\theta(\mathbf{x}) \approx f(\mathbf{x}),
\]
enabling effective estimation and cancellation of PIM interference in the receiver.

\section{Methodology}

In this section, we present the proposed methodology for efficient passive intermodulation cancellation. We first introduce the key convolutional building blocks—standard convolution, depthwise separable convolution, and dilated convolution—and then describe the overall model architecture, training strategies, and optimization techniques. Together, these components enable lightweight yet effective neural network designs suitable for resource-constrained wireless communication systems.

\subsection{Convolutional Building Blocks}

\begin{figure}[h]
	\centering
	\includegraphics[width=0.4\textwidth]{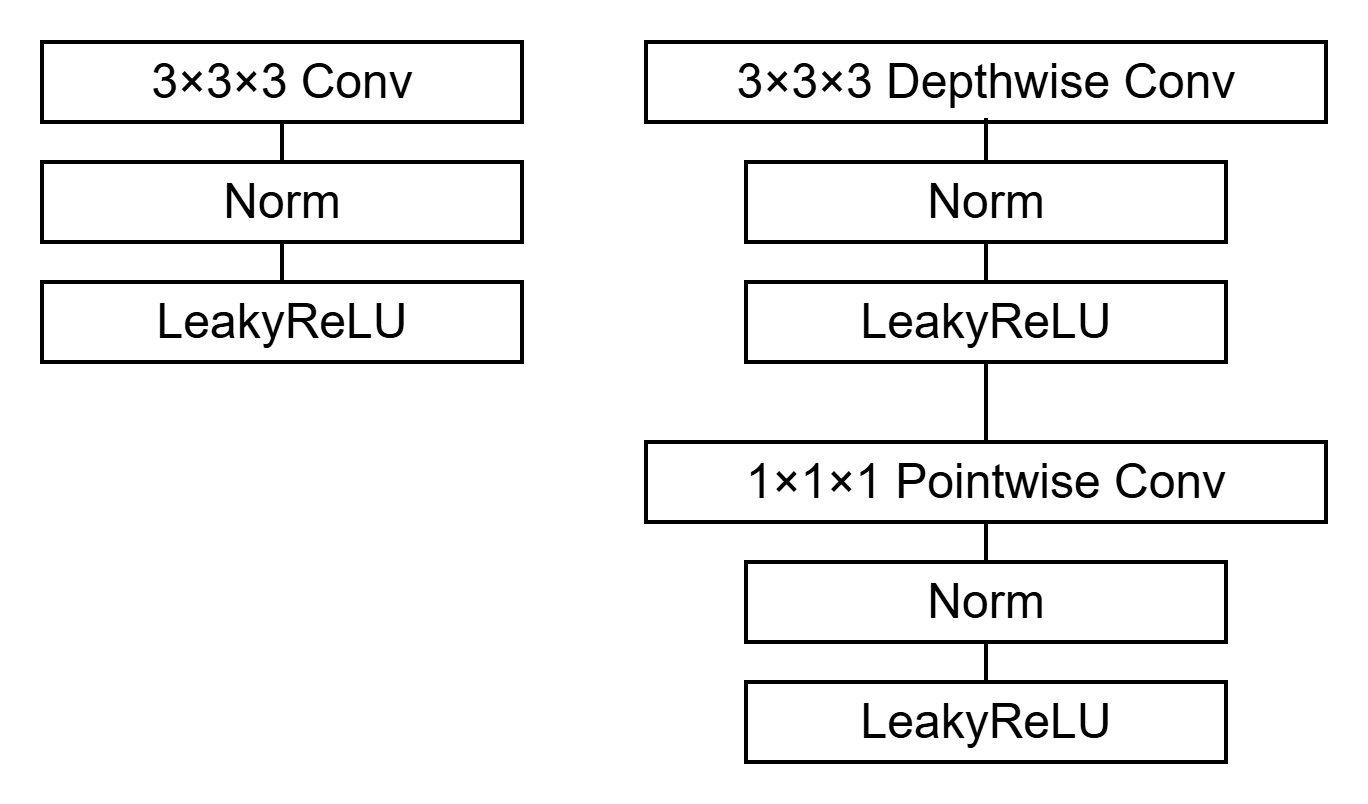}
	\caption{Left: Standard convolution with norm and LeakyReLU. Right: Depthwise Separable convolutions with norm and LeakyReLU.}
	\label{Depthwise Separable Convolutions}
\end{figure}

\subsubsection{Standard Convolution}

\

\begin{figure}[h]
    \centering
    \includegraphics[width=0.25\linewidth]{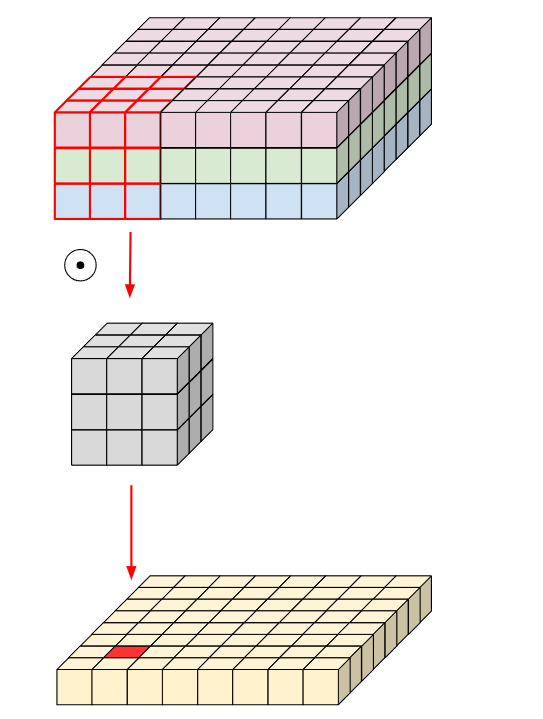}
    \caption{Standard Convolution}
    \label{Standard Convolution}
\end{figure}

\textbf{Standard convolution} (Fig. \ref{Standard Convolution} \citep{bendersky2018depthwise}) in 3D applies a kernel to the input to produce the output.

\begin{itemize}
    \item \textbf{Kernel}: A tensor of size $ (k, k, k, C_{in}, C_{out}) $, where $ (k, k, k) $ is the kernel size.
    \item \textbf{Output}: A tensor of size $ (H, W, Z, C_{out}) $.
    \item \textbf{Parameter Count}: The total number of parameters is $ k^3 \cdot C_{in} \cdot C_{out} $.
\end{itemize}

\subsubsection{Depthwise Convolution}

\begin{figure}[h]
    \centering
    \begin{subfigure}[b]{0.48\linewidth}
        \centering
        \includegraphics[width=\linewidth]{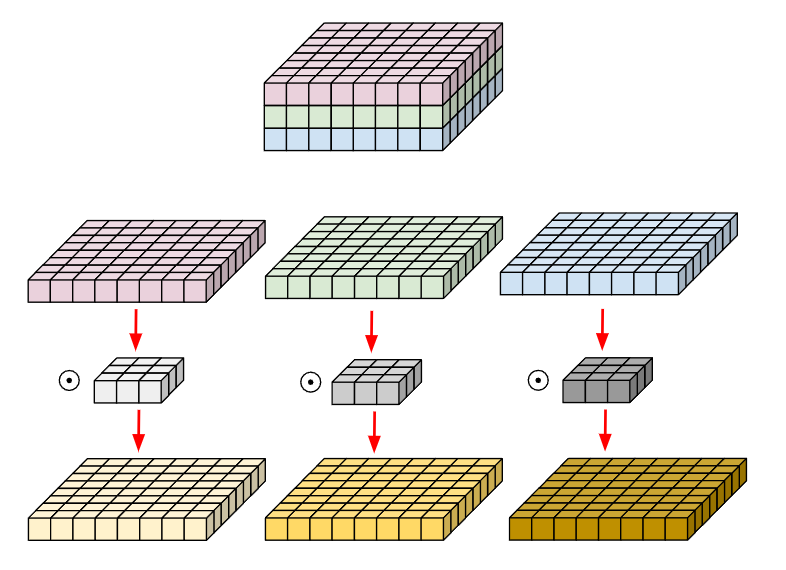}
        \caption{Depthwise Convolution}
        \label{fig:depthwise}
    \end{subfigure}
    \hfill
    \begin{subfigure}[b]{0.48\linewidth}
        \centering
        \includegraphics[width=\linewidth]{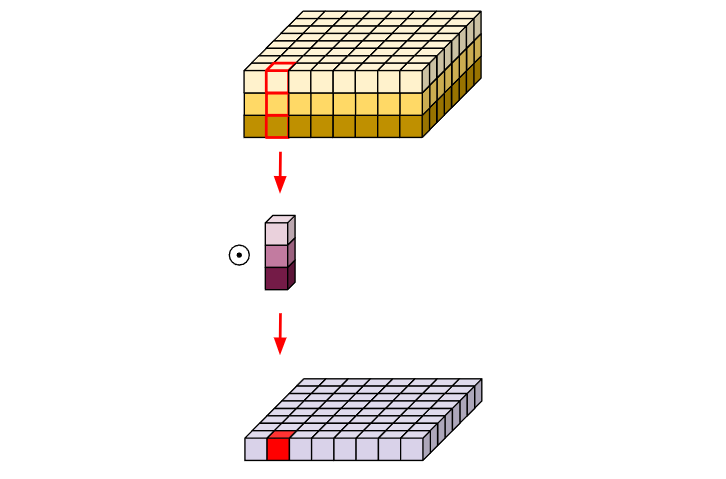}
        \caption{Pointwise Convolution}
        \label{fig:pointwise}
    \end{subfigure}
    \caption{Depthwise and Pointwise Convolution}
    \label{fig:depth_point}
\end{figure}

\textbf{Depthwise convolution} (Fig. \ref{fig:depthwise} \citep{bendersky2018depthwise}) applies a single filter to each input channel independently.

\begin{itemize}
    \item \textbf{Kernel}: A tensor of size $ (k, k, k, 1, C_{in}) $, where $ k $ is the kernel size.
    \item \textbf{Output}: A tensor of size $ (H, W, Z, C_{in}) $.
    \item \textbf{Parameter Count}: The total number of parameters is $ k^3 \cdot C_{in} $.
\end{itemize}

\subsection{Pointwise Convolution}

\textbf{Pointwise convolution} (Fig. \ref{fig:pointwise} \citep{bendersky2018depthwise}) uses a $ 1 \times 1 $ kernel to combine the outputs of the depthwise convolution.

\begin{itemize}
    \item \textbf{Kernel}: A tensor of size $ (1, 1, 1, C_{in}, C_{out}) $.
    \item \textbf{Output}: A tensor of size $ (H, W, Z, C_{out}) $.
    \item \textbf{Parameter Count}: The total number of parameters is $C_{in} \cdot C_{out} $.
\end{itemize}

\subsubsection{Dilated Convolution}

\

Dilated Convolution, also known as atrous convolution, is a powerful technique in convolutional neural networks (CNNs) designed to expand the receptive field of filters without increasing the number of parameters or computational cost. Unlike standard convolution, which processes input data in a contiguous manner, dilated convolution introduces gaps (or "holes") between the filter's kernel elements, controlled by a parameter called the dilation rate.

First, consider a one-dimensional signal. The output of dilated convolution is $y[i]$, the input is $x[i]$, the filter with size $K$ is $\omega[k]$, and the rate $=r$. It is defined as:

$$
y[k]=\sum_{k=1}^K x[i+r \cdot k] \cdot \omega[k]
$$

Using the above two types of dilated convolution, the performance of dilated convolution on two-dimensional signals (images) is as follows:

Upper branch: [Downsample by $1 / 2$ ] $\rightarrow$ [Convolution] $\rightarrow$ [Upsample by $2 x$ ]. Essentially, this only performs convolution on $1 / 4$ of the original image.

Lower branch: [Use dilated convolution with rate $=2$ ]. This approach achieves responses across the entire image, yielding better results.

Both operations have the same parameters and receptive fields, but the latter performs better.

Dilated convolution can expand the receptive field. A rate $=r$ introduces $r-1$ zeros, effectively expanding the receptive field from $[k \times k]$ to $[k+(k-1)(r-1)]$, without increasing parameters or computational cost. In DCNNs, it is common to mix dilated convolutions to compute the final DCNN responses at high resolution (understood as dense sampling).

\subsubsection{Depthwise Separable Dilated Convolution}

\

Depthwise separable convolution \citep{howard2017mobilenets} consists of two parts: a \textbf{depthwise convolution} and a \textbf{pointwise convolution}. Depthwise convolution integrates information within each channel, while pointwise convolution fuses information across channels. To enhance the model’s expressive capacity, we insert an activation function between the depthwise convolution and the pointwise convolution.

\begin{itemize}
    \item \textbf{Depthwise Convolution with Dilation}: Applies a single filter to each input channel with a dilation factor to capture long-range dependencies.
    \item \textbf{Pointwise Convolution}: Combines the outputs of the depthwise convolution using a $ 1 \times 1 $ kernel.
    \item \textbf{Parameter Count}: The total number of parameters remains $ k^3 \cdot C_{in} + C_{in} \cdot C_{out} $.
\end{itemize}

The ratio of computational complexity to parameter count for depthwise separable convolution is given by:

$$
\frac{k^3 \cdot C_{in} + C_{in} \cdot C_{out}}{k^3 \cdot C_{in} \cdot C_{out}} = \frac{k^3 + C_{out}}{k^3 \cdot C_{out}}
$$

Depthwise separable convolution offers a significant advantage in reducing the number of parameters while maintaining effective feature extraction capabilities. This makes depthwise separable convolution particularly suitable for lightweight and efficient neural network designs, especially in scenarios where computational resources are limited.

Depthwise separable dilated convolution combines depthwise convolution with a dilation factor to increase the receptive field without increasing the number of parameters.

\begin{itemize}
    \item \textbf{Depthwise Convolution with Dilation}: Applies a single filter to each input channel with a dilation factor to capture long-range dependencies.
    \item \textbf{Pointwise Convolution}: Combines the outputs of the depthwise convolution using a $ 1 \times 1 $ kernel.
    \item \textbf{Parameter Count}: The total number of parameters remains $ k \cdot C_{in} + C_{in} \cdot C_{out} $.
    \item \textbf{Computational Complexity}: The total number of operations is $ T \cdot k \cdot C_{in} + T \cdot C_{in} \cdot C_{out} $.
\end{itemize}

Depthwise separable convolution offers a significant advantage in reducing the number of parameters while maintaining effective feature extraction capabilities. By decoupling the spatial and channel-wise filtering operations, it achieves this through two key steps:

\begin{itemize}

\item 1. Depthwise Convolution: Applies a single filter independently to each input channel, focusing on spatial feature extraction. This step drastically reduces the number of parameters compared to standard convolution, as it avoids the need for channel-wise interactions at this stage.

\item 2. Pointwise Convolution: Combines the outputs of the depthwise convolution using a $1 \times 1$ kernel, enabling channel-wise interactions and feature aggregation. This step introduces additional parameters but remains computationally efficient due to its simplicity.
\end{itemize}

The combination of these two steps results in a substantial reduction in both parameter count and computational complexity, as demonstrated by the ratio $\frac{k+C_{\text {out }}}{k \cdot C_{\text {out }}}$. This makes depthwise separable convolution particularly suitable for lightweight and efficient neural network designs, especially in scenarios where computational resources are limited.

Furthermore, when combined with dilation factors, depthwise separable dilated convolution extends the receptive field without significantly increasing the number of parameters. This allows the model to capture long-range dependencies in the data while maintaining its efficiency.

\subsection{Model Architecture}

To address the requirements of different tasks, we design two distinct neural architectures. Our design choices (number of layers, activation functions, etc.) are informed both by empirical experiments and by insights from prior work.

\subsubsection{Architectures for Static and Dynamic Data}

\paragraph{Static Data Tasks}  
For static (time‐invariant) PIM cancellation tasks, the model has the following structure:

\begin{itemize}
  \item Two convolutional neural network (CNN) layers for feature extraction.
  \item A lookup table (LUT) interpolation layer as a nonlinear mapping kernel, followed by a ReLU activation.
  \item Two more CNN layers to further refine the representation and output.
\end{itemize}

This structure is designed to capture both **local spatial‐frequency correlations** (via the CNNs) and **nonlinear distortions** (via the LUT + ReLU). The LUT allows flexible nonlinear mapping without committing to a large fully connected network, which helps reduce parameter count and overfitting risk.

\paragraph{Dynamic Data Tasks}  
When signal features vary over time (or across transmit/receive configurations), temporal dependencies and more complex nonlinear behavior must be modeled. For this, we employ:

\begin{itemize}
  \item Two CNN layers for initial feature extraction (capturing local spectral/spatial structure).
  \item Three fully connected (FC) layers, each with a Sigmoid activation, to allow flexible nonlinear mixing and adaptation to dynamic variation.
  \item Finally, two more CNN layers to refine the output and restore localized structure.
\end{itemize}

\subsubsection{Support from Prior Literature and Motivations}

\paragraph{Why LUT + ReLU in Static Tasks}  
Lookup‐table or piecewise interpolation methods have been used historically in PIM modeling and certain hardware‐based nonlinear systems to approximate static nonlinear mappings with relatively low complexity. While LUTs are less common in modern deep learning for large dynamic datasets, they remain effective when the nonlinearity is relatively “fixed” and does not vary dramatically over time or inputs. Using a LUT layer followed by ReLU ensures that the nonlinearity is expressive but also constrained (by the LUT’s resolution and the ReLU behavior) to avoid unbounded or pathological responses.

\paragraph{Why Fully Connected + Sigmoid in Dynamic Tasks}  
Fully connected layers are powerful in modeling nonlinear interactions, particularly when inputs (features extracted by CNNs) already incorporate spatial or frequency structure. The use of Sigmoid activation in the FC layers is motivated by two considerations:

\begin{itemize}
  \item Sigmoid provides smooth, bounded nonlinearities. Because dynamic tasks often involve variations (e.g. due to environmental or hardware changes), smooth transitions in output behavior can help avoid abrupt changes or artifacts.
  \item In some literature, sigmoid‐like activations are still preferred in shallower or smaller dynamic networks when modeling nonlinearities, gating, or modulation effects, particularly when the outputs must lie within certain ranges. While ReLU is more popular in deep networks, Sigmoid remains relevant in contexts where outputs are interpreted as “activations” or weights rather than raw unbounded predictions.  
\end{itemize}

However, Sigmoid also has known limitations: vanishing gradient problems \citep{krizhevsky2012imagenet} (for extreme inputs) and slower convergence compared with piecewise linear activations (like ReLU). Prior studies \citep{mesran2024investigating} have observed that in many image classification tasks, ReLU tends to lead to \emph{faster convergence} and \emph{better accuracy} than sigmoid, especially as the network depth or model size increases. 
We mitigate the limitations via using only a moderate number of FC layers (three) and combining them with CNN layers above and below.

\paragraph{Choice of Activation Functions: ReLU vs Sigmoid}

\begin{itemize}
  \item \textbf{ReLU} is used following LUT in the static model to gain its advantages: non‐saturating gradient (for positive inputs), computational simplicity, and empirical success in deep CNNs. In many recent works, ReLU (or its variants) outperforms sigmoid/tanh in terms of convergence speed and training stability. :contentReference[oaicite:1]{index=1}  
  \item \textbf{Sigmoid} is adopted in FC layers of the dynamic model where bounded, smooth nonlinear transformations are desirable. While literature often cautions about vanishing gradients when using sigmoid, in contexts where the network depth is modest and input activations are normalized, it remains viable.  
\end{itemize}

\subsubsection{Summary}

In sum, our architecture choices—CNN + LUT + ReLU for static, versus CNN + FC + Sigmoid + CNN for dynamic—are guided by balancing expressiveness, flexibility, parameter efficiency, and stability. These designs draw from empirical practices in signal processing and deep learning, and we found via preliminary experiments that they yield strong PIM cancellation performance across both static and dynamic conditions.

\subsection{Learning Rate Scheduler}

\

In deep learning, the choice of learning rate schedule strongly influences both convergence speed and final model performance. We adopt a cyclical learning rate (CLR) scheduler \citep{smith2017cyclical}, which dynamically varies the learning rate between a pre-defined minimum and maximum according to a fixed period. By allowing the learning rate to fluctuate during training, CLR encourages the model to explore the parameter space more effectively and helps it escape shallow local minima without the need for manual restarts or extensive tuning.

Experimental studies have shown that cyclical learning rates can accelerate convergence and improve generalization compared with static or monotonically decreasing learning rate schedules \citep{smith2017cyclical, smith2019super, gao2021empirical}. In particular, Smith \citep{smith2017cyclical} demonstrated that cyclic policies could achieve comparable or even superior results to longer training with constant or exponentially decaying learning rates, while reducing the total training time. Follow-up work \citep{smith2019super} further confirmed that CLR benefits a variety of architectures and tasks, including image classification and semantic segmentation, by enabling the optimizer to traverse saddle points and escape poor local minima.

The underlying principle of CLR is that larger learning rates in the early stages of training allow the model to explore a wide range of solutions quickly, while smaller learning rates later help fine-tune the parameters toward high-quality minima \citep{li2019cyclical}. Moreover, empirical analyses suggest that cyclic schedules can implicitly regularize the model by preventing overfitting and encouraging exploration of flatter regions in the loss landscape \citep{keskar2017large}.

In our experiments, we found that using CLR not only accelerates convergence but also leads to better performance on the test set. The model reaches comparable or lower loss values than conventional static schedules in fewer iterations, demonstrating the efficiency and effectiveness of this approach.

\subsection{Optimizer}

We employ the Adam optimizer \citep{kingma2014adam} with a weight decay of 0.01. Weight decay acts as a regularization technique by penalizing large parameter values, which mitigates overfitting and enhances the model's generalization ability \citep{krogh1992simple}. Adam combines the benefits of momentum-based optimization and adaptive learning rates, making it particularly effective for training deep neural networks across diverse tasks \citep{kingma2014adam}.

Compared to traditional Stochastic Gradient Descent (SGD) \citep{bottou2010large}, Adam offers several key advantages:

Adaptive Learning Rates: Adam dynamically adjusts the learning rate for each parameter based on estimates of the first moment (mean) and second moment (uncentered variance) of the gradients. This allows the optimizer to scale updates for parameters with high gradient variability, stabilizing training, while accelerating convergence for parameters with lower gradient variance. This mechanism is particularly useful for heterogeneous networks with layers of differing sensitivity and scale, such as deep CNNs or recurrent networks \citep{kingma2014adam, ruder2016overview}.

Reduced Hyperparameter Sensitivity: Unlike SGD, which often requires careful tuning of a global learning rate and momentum terms, Adam’s adaptive mechanism reduces the need for extensive manual hyperparameter optimization. This simplifies the training process and allows practitioners to achieve robust convergence across a range of tasks without exhaustive trial-and-error \citep{goodfellow2016deep}.

Momentum Integration: Adam leverages exponentially weighted moving averages of past gradients, similar to momentum methods, to accelerate convergence and dampen oscillations in the parameter space. This is particularly beneficial in non-convex optimization landscapes, where sharp valleys and plateaus can slow convergence \citep{sutskever2013importance}.

Empirical Effectiveness: Numerous studies have demonstrated Adam’s superiority over SGD in terms of convergence speed and final performance, especially in deep or complex architectures \citep{kingma2014adam, reddi2019convergence}. It has been widely adopted in computer vision, natural language processing, and reinforcement learning tasks due to its reliability and efficiency.

Variants and Extensions: Over the years, several extensions of Adam, such as AdamW \citep{loshchilov2017decoupled}, have addressed potential shortcomings like the improper coupling of weight decay and adaptive learning rates. Using decoupled weight decay further improves generalization without compromising Adam’s convergence properties.

In our experiments, Adam with weight decay effectively balances convergence speed and test-set performance, enabling stable and efficient training across both static and dynamic data tasks.

\subsection{Boundary Effect Mitigation}

Convolutional neural networks are susceptible to boundary effects—artifacts that occur near the edges of the input data due to the lack of sufficient context for the convolutional kernels. This phenomenon is well-documented in both image processing and sequence modeling tasks, where the receptive field of a filter extends beyond the available input at the boundaries, resulting in incomplete or biased feature representations \citep{yu2015multi, chen2018deeplab}. Such effects can degrade model performance, particularly in applications requiring dense or pixel-wise predictions, such as semantic segmentation \citep{long2015fully} or time-series forecasting.

Several strategies have been proposed to mitigate boundary effects. Common approaches include padding, where zeros, reflections, or replicated values are added to the input to artificially extend its size, and truncation, which ignores the predictions near boundaries to avoid unreliable outputs \citep{gu2019recent}. In our work, we adopt a truncation-based strategy: during training, we exclude the beginning and end segments of the output sequence from loss computation. By focusing on the central, non-boundary regions, the model learns features that are less influenced by edge artifacts, improving generalization and robustness.

Additionally, boundary effects can be exacerbated in networks with large receptive fields or high dilation rates, where the convolution kernel may span a significant portion of the input \citep{yu2015multi}. Our truncation approach effectively mitigates these risks, ensuring that long-range dependencies captured by dilated convolutions do not introduce misleading signals at the boundaries. Furthermore, this method is computationally efficient, as it avoids additional operations such as mirrored padding or complex boundary-aware convolutions \citep{li2020boundary}.

In practice, mitigating boundary effects not only improves model performance on central regions but also enhances overall test-set performance, particularly for datasets where edge regions contain informative patterns. By combining truncation with careful network design (e.g., using overlapping receptive fields and dilated convolutions), we achieve a balance between capturing long-range dependencies and minimizing artifacts, consistent with best practices in modern CNN architectures \citep{yu2015multi, chen2018deeplab}.

\subsection{Batch Size}

Batch size is a crucial hyperparameter in training deep learning models, as it directly affects both the convergence behavior and computational efficiency. It controls the number of training samples processed before updating model parameters and influences gradient estimation, generalization, and memory usage \citep{keskar2016large, smith2017don}.

Smaller batch sizes introduce higher stochasticity into gradient estimates, which can help the optimizer escape sharp local minima and potentially improve generalization performance \citep{smith2018disciplined}. This stochastic effect can be especially advantageous in non-convex optimization landscapes typical of deep neural networks, where numerous local minima and saddle points exist \citep{goodfellow2016deep}. Additionally, smaller batch sizes reduce memory footprint, allowing the training of larger models or higher-resolution inputs on limited hardware resources, and enable more frequent parameter updates, sometimes leading to faster convergence \citep{li2014efficient}.

Conversely, larger batch sizes provide more accurate estimates of the gradient, resulting in smoother updates and potentially more stable convergence \citep{de2017small}. They are particularly useful in scenarios where data is noisy or when the optimization landscape is relatively flat. Larger batches can also better exploit parallel processing capabilities of modern GPUs, improving hardware utilization. However, excessively large batches have been associated with poorer generalization performance, as they tend to converge to sharp minima \citep{keskar2016large}. Moreover, extremely large batch sizes can necessitate extensive memory resources and may require careful tuning of learning rates to avoid optimization difficulties.

In our experiments, we selected a batch size larger than the FFT sampling length of 1024 to balance these trade-offs. This choice ensures the integrity of frequency-domain computations while maintaining a level of stochasticity beneficial for convergence and generalization. By carefully balancing memory efficiency, gradient noise, and convergence dynamics, our approach achieves both stable and efficient training, resulting in improved test performance.

\subsection{Gradient Clipping}

Gradient clipping is a widely used technique in deep learning designed to prevent the problem of exploding gradients, which can destabilize the training process and lead to numerical divergence \citep{pascanu2013difficulty}. Exploding gradients typically occur in deep neural networks, particularly in recurrent architectures or very deep feedforward networks, where repeated multiplications during backpropagation amplify the gradients exponentially. When left unchecked, excessively large gradients can cause abrupt updates to model parameters, resulting in unstable training dynamics or even divergence \citep{bengio1994learning}.

The fundamental idea behind gradient clipping is to rescale the gradient vector when its norm exceeds a predefined threshold. Formally, if the computed gradient \(g\) satisfies \(\|g\| > \tau\), where \(\tau\) is the clipping threshold, the gradient is rescaled as:

\[
g_{\text{clipped}} = g \cdot \frac{\tau}{\|g\|}
\]

This simple operation ensures that updates remain bounded while preserving the gradient's direction, preventing the optimizer from making overly large steps that could destabilize training \citep{goodfellow2016deep}.  

Gradient clipping provides multiple advantages beyond mere stabilization. Firstly, it improves the reliability and smoothness of the optimization trajectory, enabling consistent convergence even in very deep networks \citep{pascanu2013difficulty}. Secondly, it acts as a safeguard against noisy or irregular data, which may produce sharp gradient spikes in highly non-convex loss landscapes. By limiting the magnitude of these extreme gradients, clipping reduces the risk of the optimizer overshooting minima, thereby enhancing both convergence and generalization \citep{zhang2019gradient}.  

This technique is particularly valuable for training recurrent neural networks (RNNs) and long short-term memory (LSTM) models, where gradients can accumulate over time steps and amplify exponentially \citep{pascanu2013difficulty, chung2014empirical}. In such cases, gradient clipping not only stabilizes training but also allows the use of larger learning rates, accelerating convergence without compromising model performance. Moreover, gradient clipping has been shown to be effective in transformer architectures and very deep convolutional networks, where deep or residual connections may introduce instability in gradient flow \citep{vaswani2017attention}.

In our implementation, gradient clipping is applied after computing the gradients and before updating the model parameters. By carefully selecting the threshold, we ensure that the gradients are scaled proportionally, preserving their direction while limiting extreme magnitudes. This approach not only stabilizes the training process but also improves generalization by reducing overfitting to outliers or noisy samples.

In summary, gradient clipping is a simple yet highly effective technique to enhance the stability and robustness of deep learning training. It ensures controlled parameter updates, prevents gradient explosion, and facilitates faster convergence. Its application is essential in complex architectures, recurrent models, and challenging datasets, where gradient instability is a common issue.

\section{Results}

\subsection{Average Power Error (APE)}

We evaluate the performance using the Average Power Error (APE) across all channels.  
Let $P_{i,c}^{\rm meas}$ and $P_{i,c}^{\rm ref}$ denote the measured and reference power of the $i$-th sample in channel $c$, with $N$ samples and $C$ channels in total. Then, the APE is defined as:

\begin{equation}
\text{APE} = \frac{1}{C} \sum_{c=1}^{C} \frac{1}{N} \sum_{i=1}^{N} \left| P_{i,c}^{\rm meas} - P_{i,c}^{\rm ref} \right|,
\end{equation}

or, if the power is expressed in dB:

\begin{equation}
\text{APE}_{\rm dB} = \frac{1}{C} \sum_{c=1}^{C} \frac{1}{N} \sum_{i=1}^{N} \left| 10 \log_{10} \frac{P_{i,c}^{\rm meas}}{P_{i,c}^{\rm ref}} \right|.
\end{equation}

Here, the outer sum averages over channels and the inner sum averages over all samples in each channel.

\subsection{PIM Cancellation Task for Static Scenario}

In this experiment, we assessed the performance of our model in mitigating Passive Intermodulation (PIM) signal interference in a static scenario, where the signals remain constant over time. The experiment was conducted using a system with 32 transmitting antennas and 16 receiving antennas. The model under evaluation contains 25,632 parameters, which are optimized during training to accurately predict and suppress the interference.

\begin{figure}[h]
\centering
\includegraphics[width=0.7\linewidth]{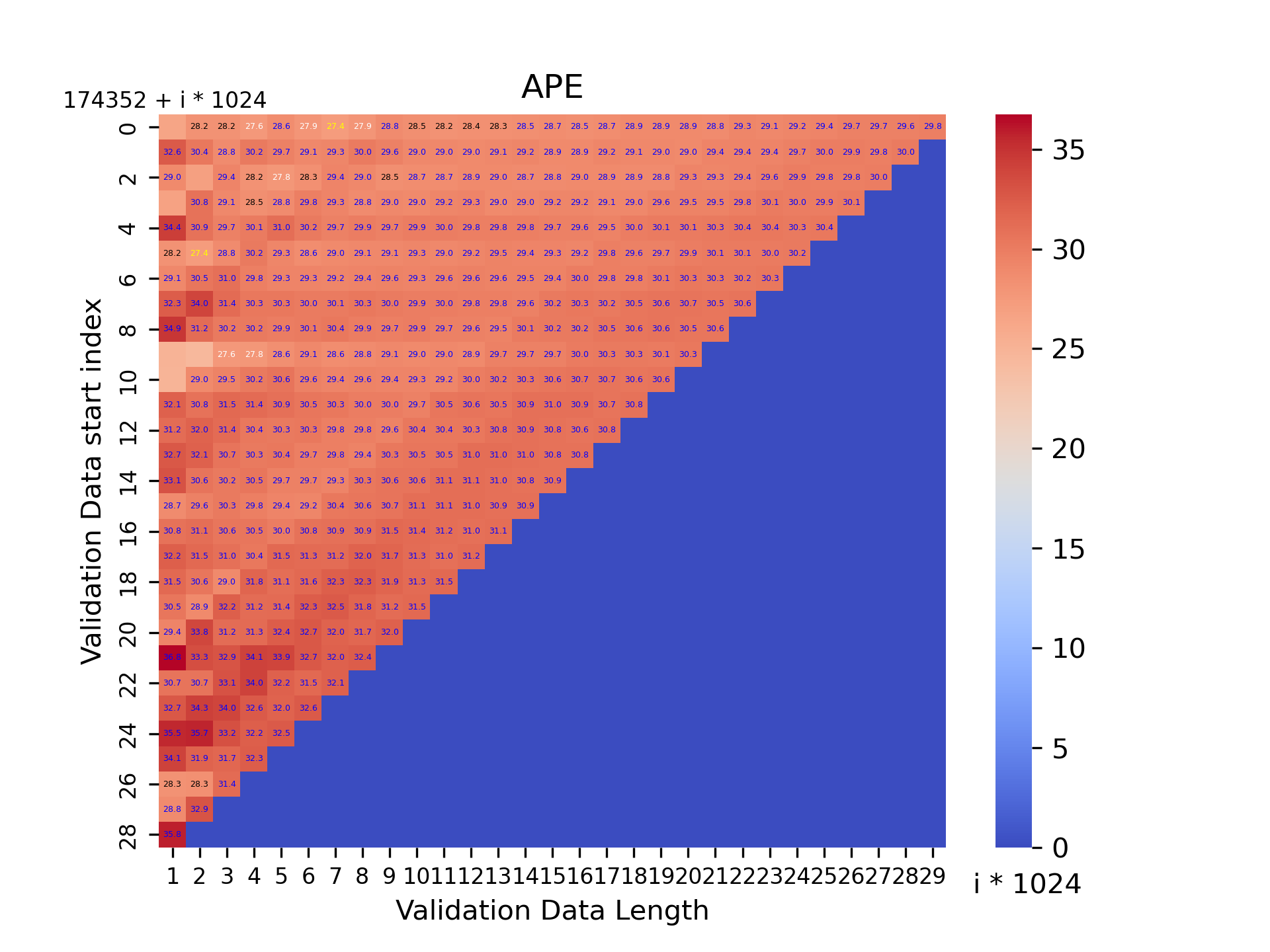}
\caption{Heatmap of APE (dB) for PIM Cancellation on Different Test Segments.}
\label{fig4}
\end{figure}

Figure~\ref{fig4} presents a heatmap depicting the model’s performance across different segments of the test data. In this heatmap:

\begin{itemize}
\item The vertical axis represents the starting position of each selected data segment.
\item The horizontal axis represents the length of each selected data segment.
\item The color encodes the Average Prediction Error (APE) in decibels (dB), with higher values indicating better performance (i.e., lower prediction error).
\end{itemize}

For instance, the point corresponding to “horizontal 29, vertical 0” reflects the model’s performance on the entire test set of 29 segments, achieving an APE of 29.8 dB.

\begin{figure}[h]
    \centering
    \begin{minipage}[b]{0.48\linewidth}
        \centering
        \includegraphics[width=\linewidth]{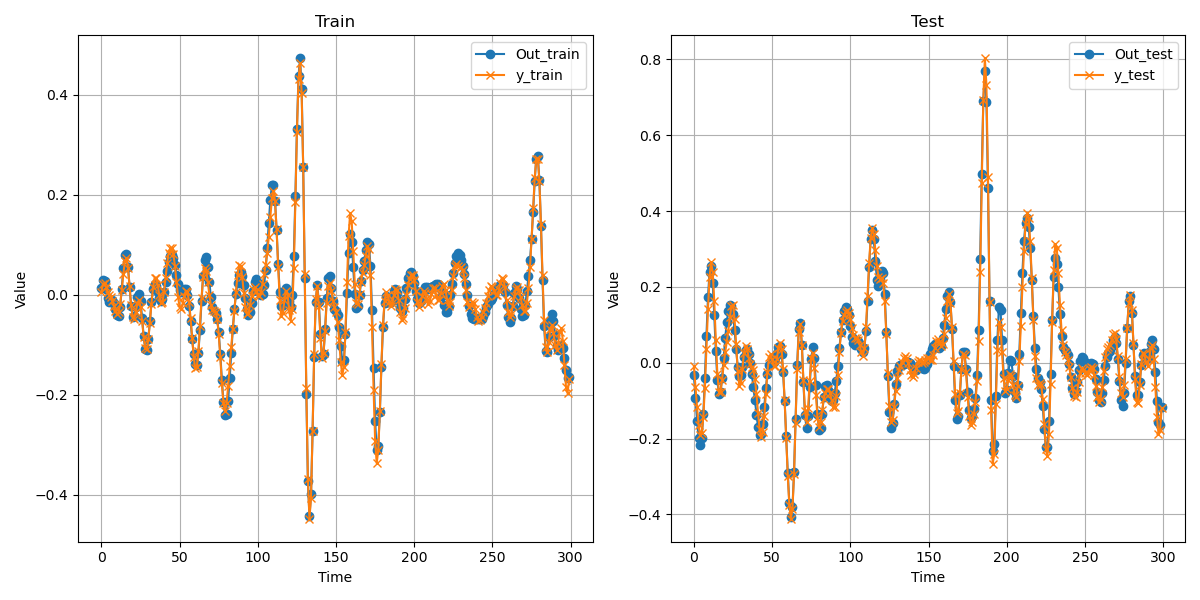}
        \caption{Time-Domain Output on Test Set: True vs. Predicted Signal}
        \label{fig10}
    \end{minipage}
    \hfill
    \begin{minipage}[b]{0.48\linewidth}
        \centering
        \includegraphics[width=\linewidth]{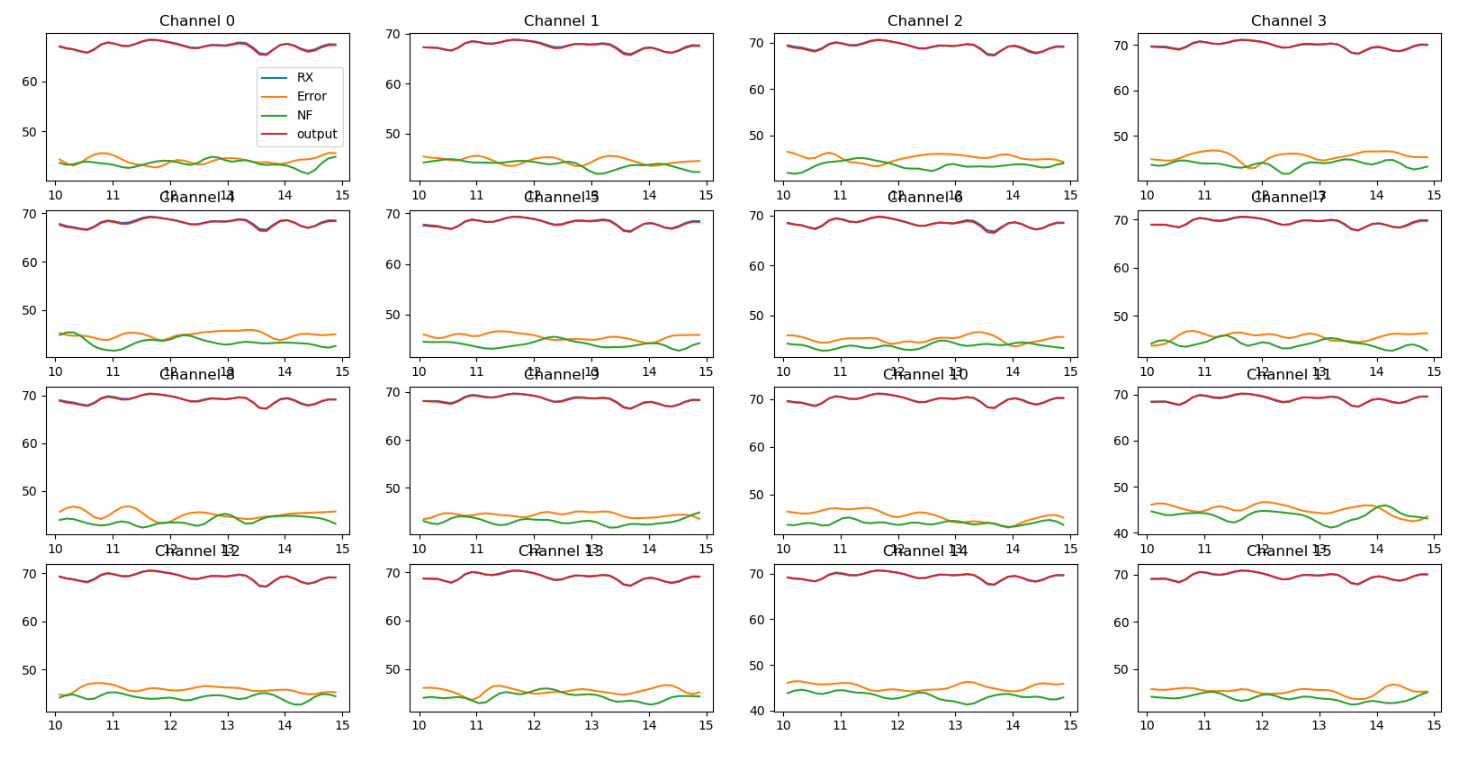}
        \caption{Frequency-Domain Output on Test Set: True vs. Predicted Signal}
        \label{fig11}
    \end{minipage}
\end{figure}

We further evaluated the model by comparing its predictions with the actual signals in both the time domain (Fig.~\ref{fig10}) and frequency domain (Fig.~\ref{fig11}). The time-domain visualization illustrates the temporal variation of the signal, whereas the frequency-domain visualization represents the signal’s spectral energy distribution. The closer the predicted waveform aligns with the true signal, the better the model’s predictive accuracy.

During training, the model achieved an APE of 44 dB on the training set. On test sets of varying lengths, the model achieved an APE of 29.8 dB on the 30k-length set and 30 dB on the 29k-length set, demonstrating consistent and robust performance across different data sizes.

To create a lighter and faster model, we reduced its size to 11,856 parameters. This lightweight model consists of:

\begin{itemize}
    \item Two convolution layers for extracting patterns from the input.
    \item Two fully connected layers with Sigmoid activation to combine information in a nonlinear way.
    \item Two additional convolution layers to refine features.
\end{itemize}

\begin{figure}[h]
\centering
\includegraphics[width=0.7\linewidth]{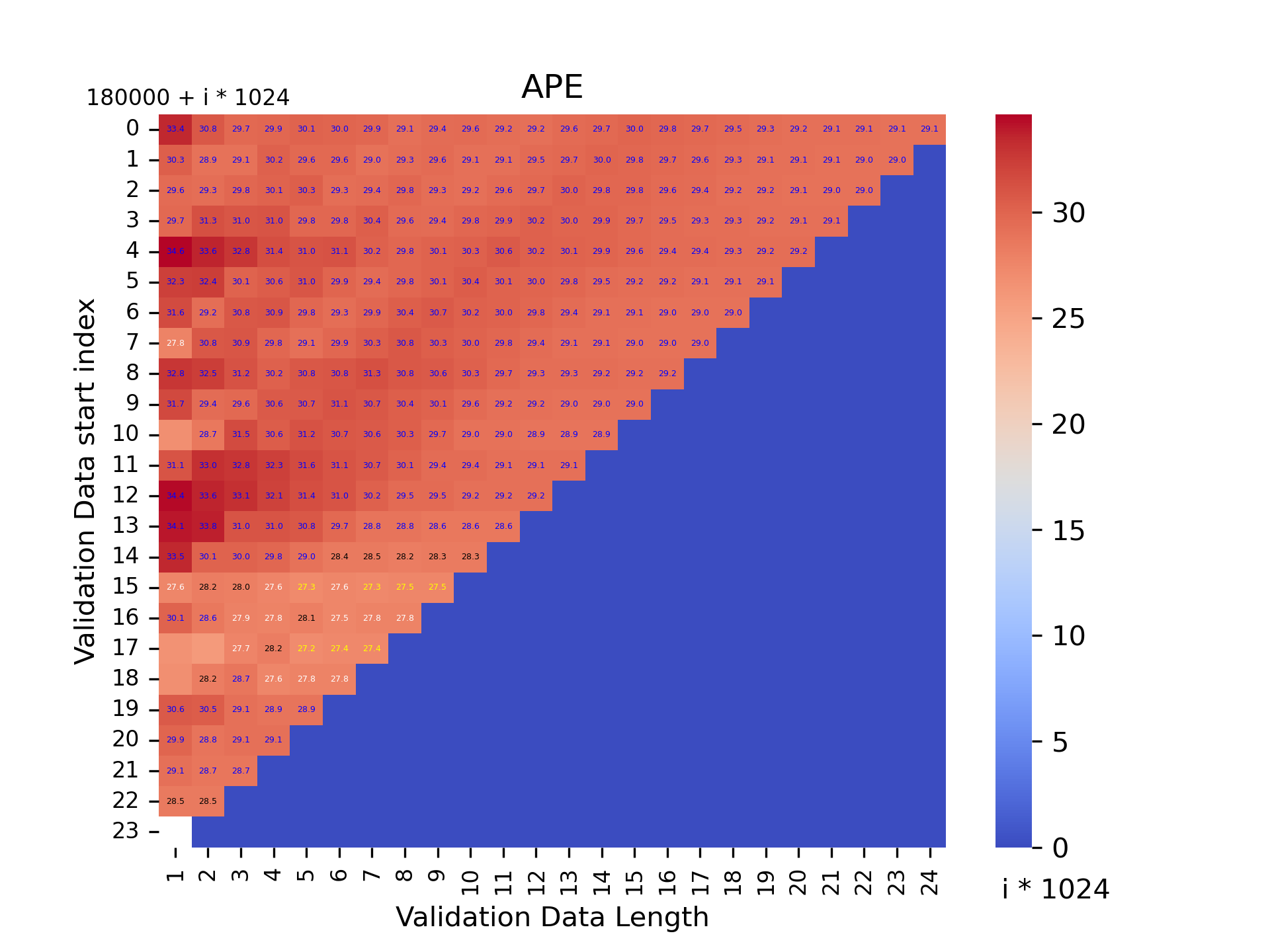}
\caption{Heatmap of APE (dB) for Lightweight Model.}
\label{fig12}
\end{figure}

\begin{figure}[h]
    \centering
    \begin{minipage}[b]{0.48\linewidth}
        \centering
        \includegraphics[width=\linewidth]{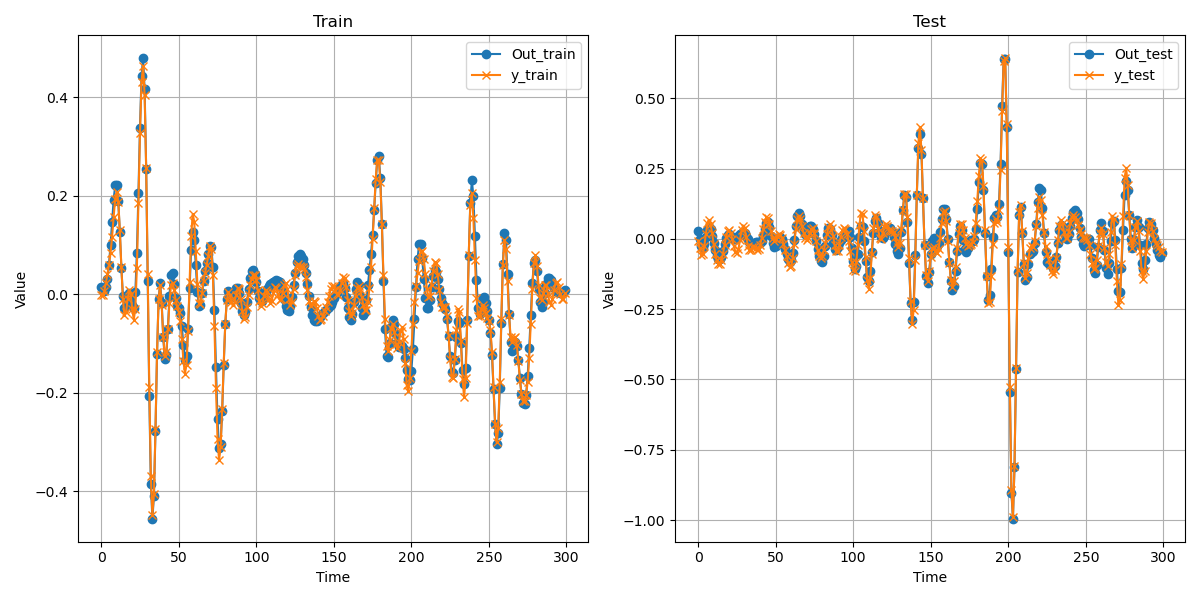}
        \caption{Time-Domain Output: True vs. Predicted Signal}
        \label{fig13}
    \end{minipage}
    \hfill
    \begin{minipage}[b]{0.48\linewidth}
        \centering
        \includegraphics[width=\linewidth]{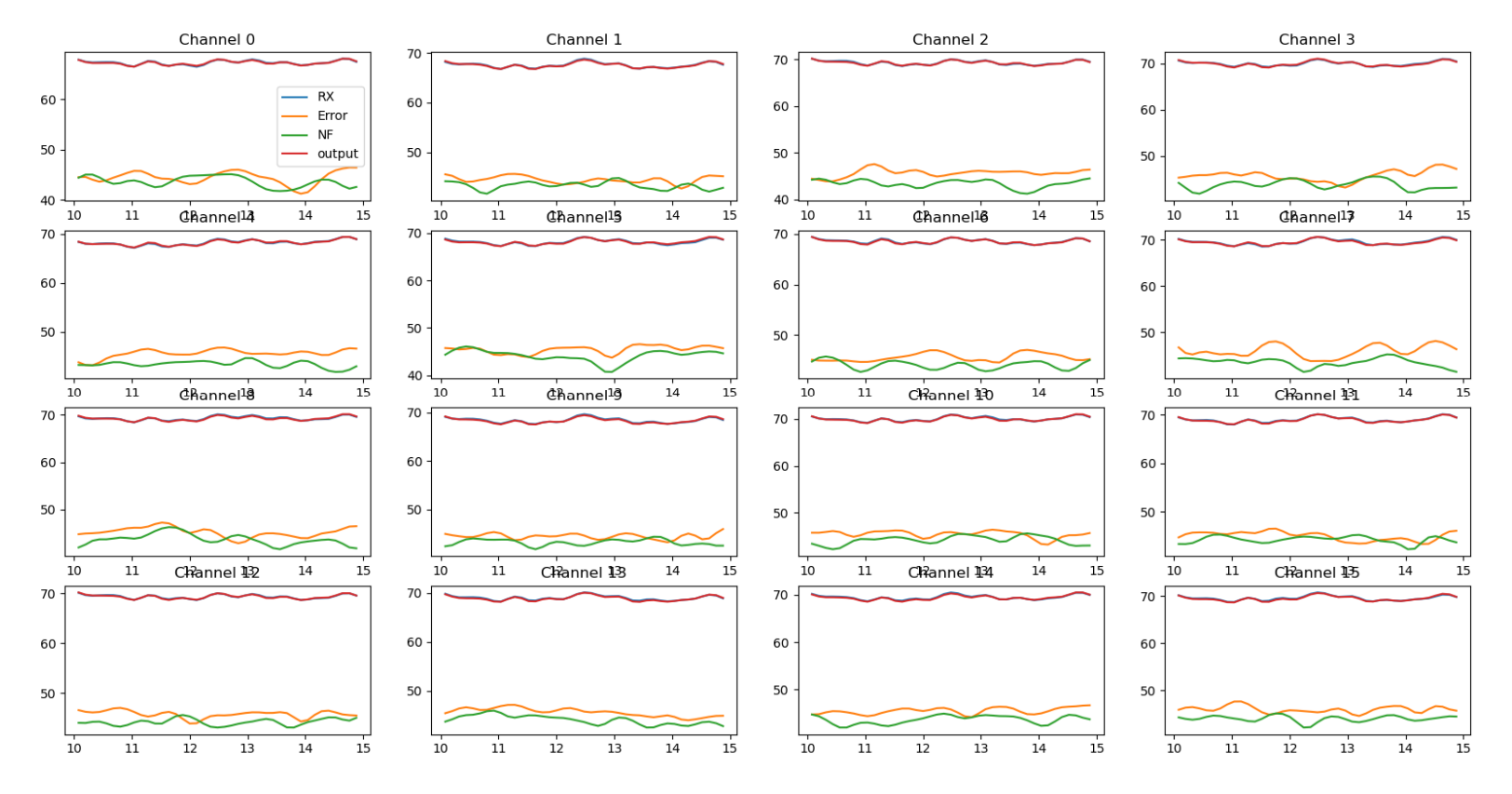}
        \caption{Frequency-Domain Output: True vs. Predicted Signal}
        \label{fig14}
    \end{minipage}
\end{figure}

Even with fewer parameters, the model still performs well. On test segments of length 25k, it achieves an APE of 29 dB, and on segments of length 15k, an APE of 30 dB. Figures~\ref{fig12}-\ref{fig14} show heatmaps and comparisons in time and frequency domains for this lightweight model.

\subsection{PIM Cancellation Task for Dynamic Scenario}

In real-world systems, signals often change over time due to equipment behavior or environmental factors. In these dynamic scenarios, PIM interference also changes, making cancellation more challenging. Our model is designed to adapt to these changes in real time.

\begin{figure}[h]
    \centering
    \begin{subfigure}[b]{0.3\linewidth}
        \centering
        \includegraphics[width=\linewidth]{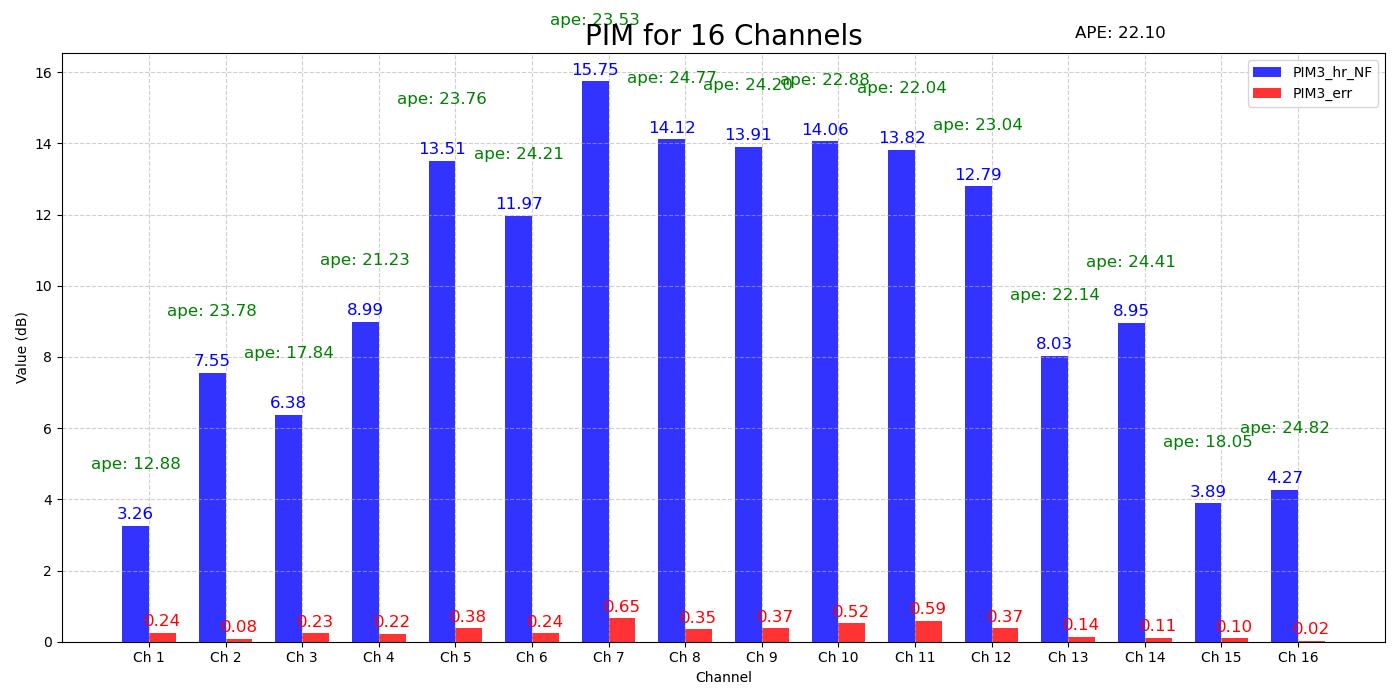}
        \caption{Dataset 1}
        \label{fig5}
    \end{subfigure}
    \hfill
    \begin{subfigure}[b]{0.3\linewidth}
        \centering
        \includegraphics[width=\linewidth]{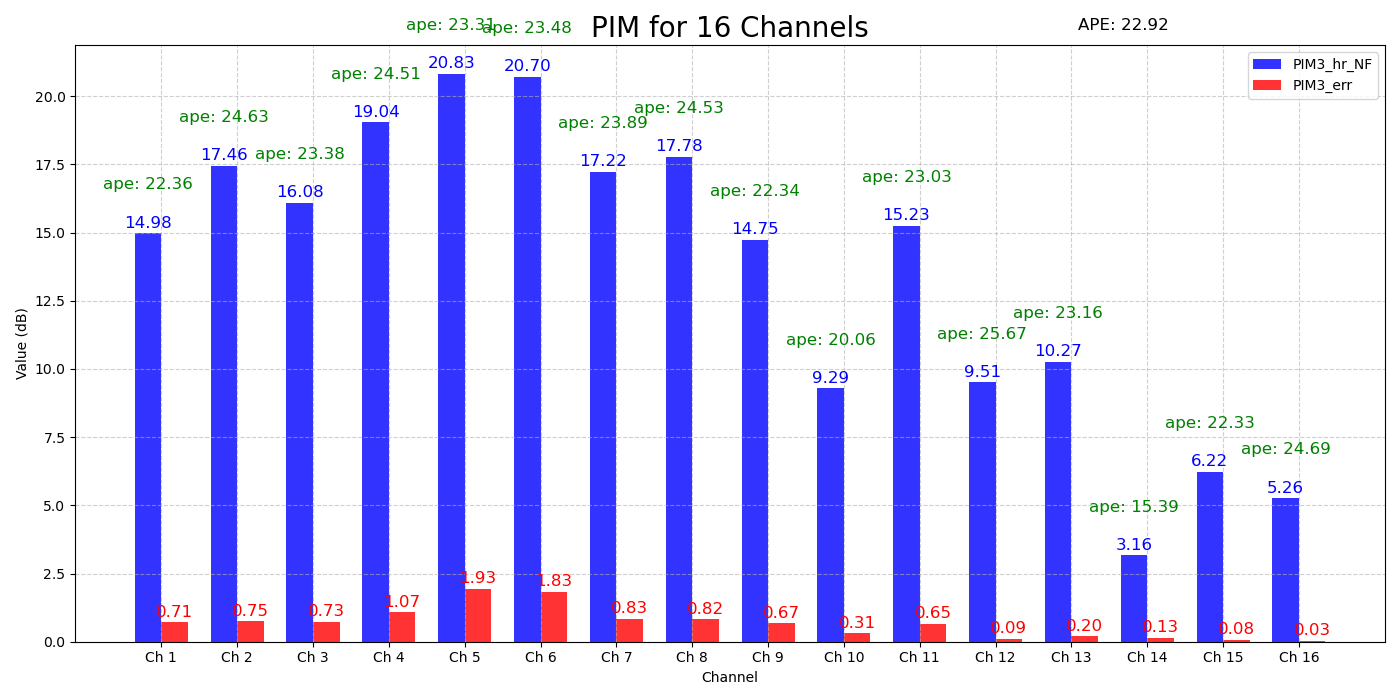}
        \caption{Dataset 2}
        \label{fig6}
    \end{subfigure}
    \hfill
    \begin{subfigure}[b]{0.3\linewidth}
        \centering
        \includegraphics[width=\linewidth]{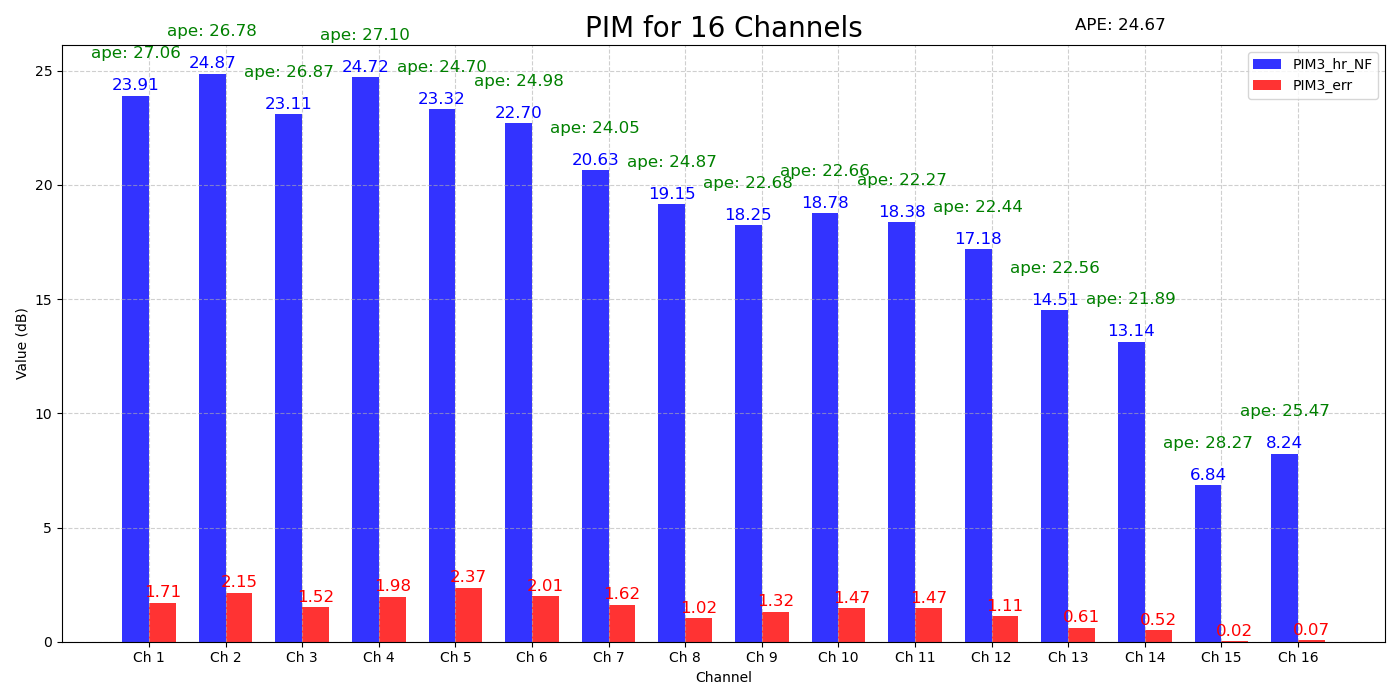}
        \caption{Dataset 3}
        \label{fig7}
    \end{subfigure}

    \vspace{0.1cm} % 行间距，可调节

     % 第二行：两图居中
    \hspace*{\fill}%
    \begin{subfigure}[b]{0.3\linewidth}
        \centering
        \includegraphics[width=\linewidth]{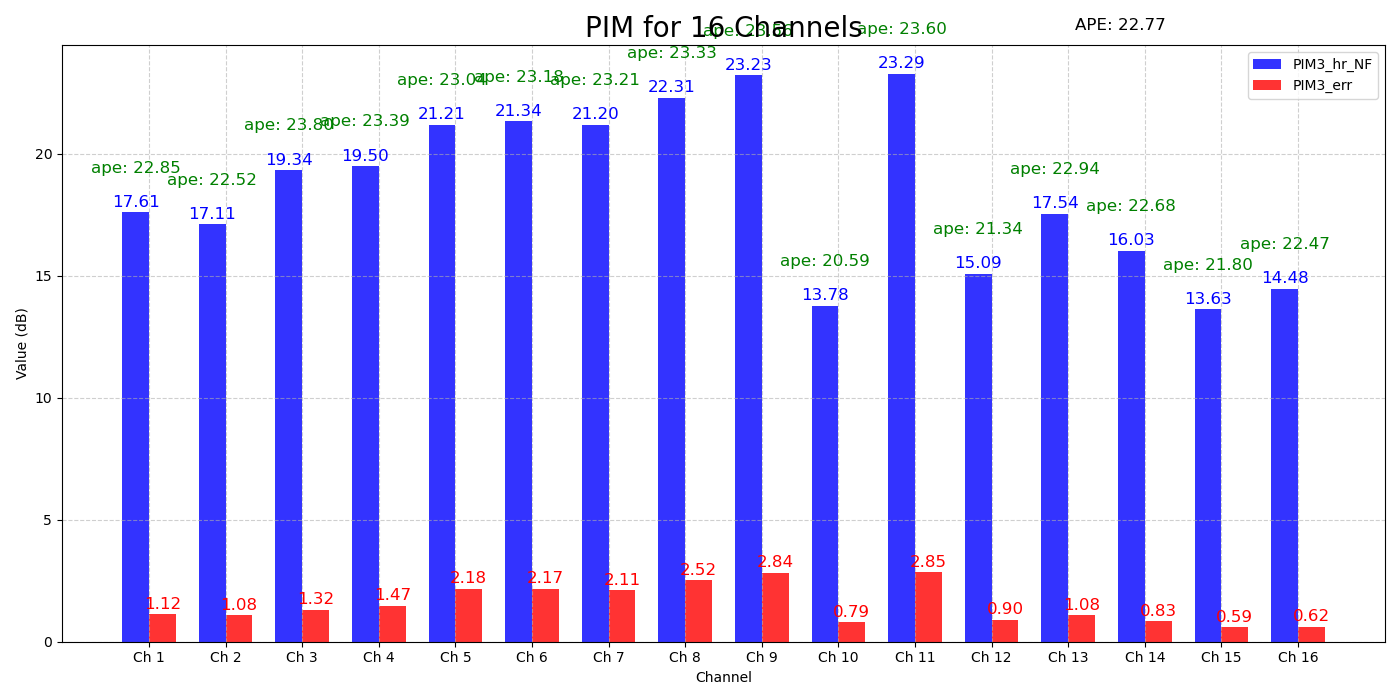}
        \caption{Dataset 4}
        \label{fig8}
    \end{subfigure}
    \hfill
    \begin{subfigure}[b]{0.3\linewidth}
        \centering
        \includegraphics[width=\linewidth]{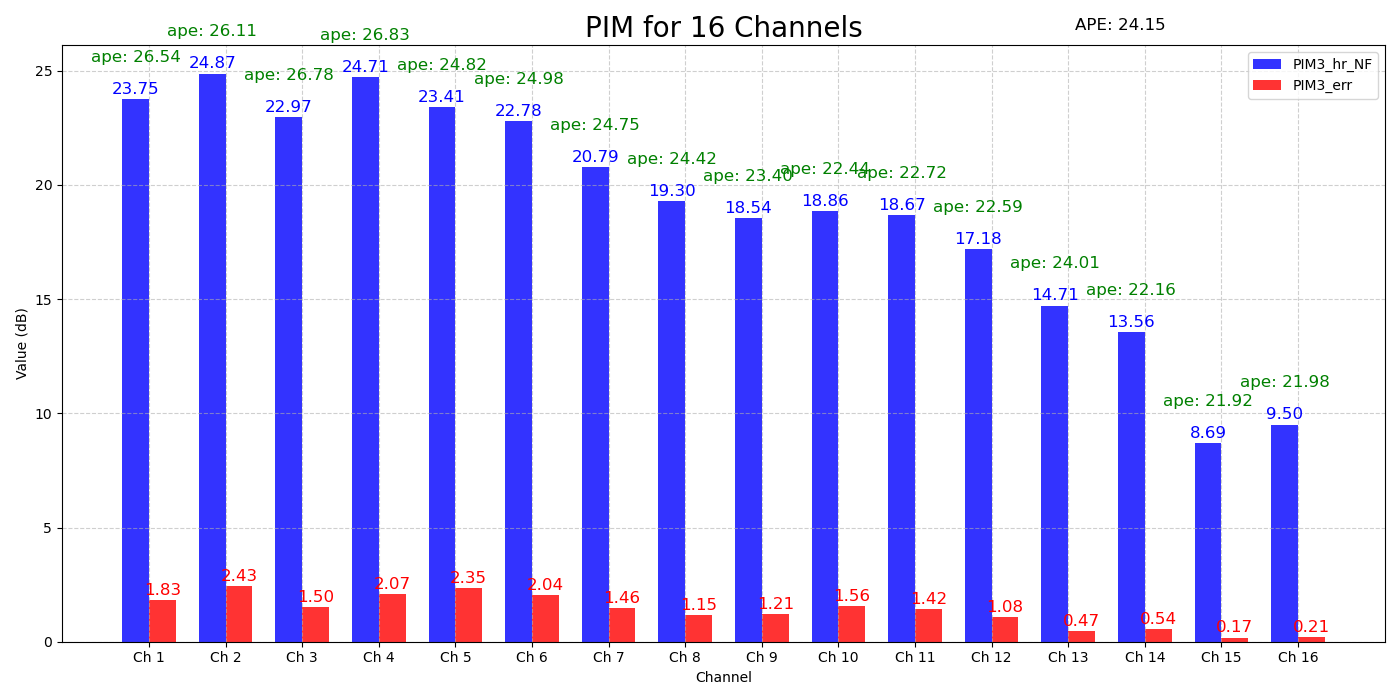}
        \caption{Dataset 5}
        \label{fig9}
    \end{subfigure}
    \hspace*{\fill}

    \caption{PIM Magnitude and Cancellation Across Channels for Five Datasets.}
    \label{fig:PIM_all}
\end{figure}

We tested the model on five different dynamic datasets. Figures~\ref{fig5}-\ref{fig9} show the measured PIM magnitude (the strength of the interference) and how much our method reduces it in different channels.  

\begin{itemize}
    \item Each bar in the charts represents the interference level in one channel.
    \item The difference between the original and the reduced PIM indicates the effectiveness of the cancellation.
\end{itemize}

The results show that our method significantly reduces PIM across all channels, and is especially effective in channels with stronger interference. This demonstrates that the model can adapt to changing conditions and maintain high performance.

\subsection{Discussion and Summary}

In this work, we developed a deep learning-based model to mitigate Passive Intermodulation (PIM) interference in MIMO wireless systems. The model was evaluated under both static and dynamic scenarios, demonstrating robust performance and efficiency. Key findings are as follows:

\begin{itemize}
    \item \textbf{Static Scenario Performance:} In static settings, where the channel conditions and signal characteristics remain constant over time, the model effectively captures deterministic PIM patterns using a relatively small number of parameters. Despite its compact size, the lightweight model achieves near-identical performance compared to a larger model, with Average Prediction Error (APE) consistently above 29 dB, indicating highly accurate PIM estimation and cancellation.
    
    \item \textbf{Dynamic Scenario Adaptability:} In dynamic environments, where signal characteristics change over time due to channel variations or mobility, the model successfully adapts to these changes, maintaining effective PIM suppression across multiple channels. The results demonstrate that the model generalizes well to unseen channel conditions, a critical property for real-world deployment.
    
    \item \textbf{Parameter Efficiency:} The lightweight model requires significantly fewer parameters than the larger counterpart, highlighting that accurate PIM cancellation does not necessarily require large model capacity. This efficiency is particularly valuable for deployment in base stations or mobile devices with limited computational resources.
    
    \item \textbf{Cross-Channel Performance:} As shown in Table~\ref{tab:ape_dynamic}, the model consistently reduces PIM interference across multiple channels, demonstrating its ability to handle multi-channel interactions and nonlinearities in passive components.
\end{itemize}

These findings suggest several broader implications and future directions:

\begin{enumerate}
    \item \textbf{Practical Deployment:} The combination of high accuracy and low model complexity makes the approach suitable for real-time implementation in 5G and future 6G base stations, where both computational efficiency and interference mitigation are critical.
    
    \item \textbf{Extension to Wider Scenarios:} While the current experiments focus on a limited set of frequency bands and antenna configurations, the model can potentially be extended to multi-band, multi-cell, and massive MIMO systems. Future work could also explore joint mitigation of PIM along with other nonlinear distortions in the RF chain.
    
    \item \textbf{Integration with Adaptive Systems:} Given its ability to generalize to dynamic scenarios, the model could be integrated with adaptive beamforming or resource allocation strategies to further enhance system performance in varying traffic and environmental conditions.
    
    \item \textbf{Robustness and Explainability:} Future studies could focus on quantifying the model’s robustness against extreme nonlinearities, hardware aging, and environmental variations, as well as developing interpretability techniques to understand which signal features contribute most to PIM prediction.
\end{enumerate}

\begin{table}[h] \centering \caption{Average Prediction Error (APE) in dB for Static Scenario} \begin{tabular}{c|c|c} \hline Test Set Length & Large Model & Lightweight Model \\ \hline 30k & 29.8 & 29.5 \\ 29k & 30 & 29.8 \\ 25k & 29.2 & 29 \\ 15k & 30 & 30 \\ \hline \end{tabular} \label{tab:ape_static} \end{table}

\begin{table}[h] \centering \caption{PIM Reduction Across Channels in Dynamic Scenario (Illustrative)} \begin{tabular}{c|c|c} \hline Channel & Original PIM (dB) & After Cancellation (dB) \\ \hline 1 & 10 & 2 \\ 2 & 8 & 1 \\ 3 & 12 & 3 \\ 4 & 9 & 1 \\ 5 & 11 & 2 \\ \hline \end{tabular} \label{tab:ape_dynamic} \end{table}

In conclusion, our approach demonstrates that deep learning can provide a practical, efficient, and accurate solution for mitigating PIM interference in modern wireless communication systems. The model’s adaptability, lightweight design, and strong performance across both static and dynamic scenarios make it a promising candidate for next-generation wireless networks.

\bibliographystyle{unsrtnat}
\bibliography{bibfile}

\end{document}